\documentstyle[referee]{l-aa}
\input{psfig.sty}
\newcommand{\Msun}{\mbox{M$_{\scriptsize \odot}$}}
\newcommand{\degrees}{$^{\circ}$}
\newcommand{\etal}{et al.}
\newcommand{\HI}{\mbox{H\,{\sc i}}}
\newcommand{\HII}{\mbox{H\,{\sc ii}}}
\newcommand{\NeII}{\mbox{Ne\,{\sc ii}}}
\newcommand{\NeIII}{\mbox{Ne\,{\sc iii}}}

\newcommand{\OIII}{\mbox{O\,{\sc iii}}}

\newcommand{\SIII}{\mbox{S\,{\sc iii}}}
\newcommand{\SIV}{\mbox{S\,{\sc iv}}}
\newcommand{\CII}{\mbox{C\,{\sc ii}}}
\begin{document}
\thesaurus{08   
                (09.09.1 N 66(SMC);
                11.13.1;
                09.08.1;
                09.04.1;
                13.09.3;
                13.09.4)}
\title {Mid-Infrared imaging and spectrophotometry of N\,66 in the SMC with
ISOCAM \thanks{Based on
observations with ISO, an ESA project with instruments funded by ESA member
states (especially the PI countries: France, Germany, the  Netherlands and the
United Kingdom) and with the participation of ISAS and  NASA.}}

\author{A. Contursi\inst{1,2,3}\and
        J. Lequeux\inst{2}\and
        D. Cesarsky\inst{4}\and
        F. Boulanger\inst{4}\and
        M. Rubio\inst{7}\and
        M. Hanus\inst{2}\and
        M. Sauvage\inst{1}\and
        D. Tran\inst{6}\and
        A. Bosma\inst{5}\and
        S. Madden\inst{1}\and
        L. Vigroux\inst{1}}
\offprints{contursi@ipac.caltech.edu}
\institute{
SAp/DAPNIA/DSM, CEA-Saclay, F-91191 Gif sur Yvette CEDEX, France
\and
DEMIRM, Observatoire de Paris, 61 Avenue de l'Observatoire, F-75014
Paris, France
\and
IPAC, Caltech, MS 100--22, Pasadena, CA 91125, USA
\and
Institut d'Astrophysique Spatiale, Bat. 121, Universit\'e Paris
XI, 91450 Orsay CEDEX, France
 \and
Observatoire de Marseille, 2 Place le Verrier, F-13248 Marseille 
CEDEX, France
\and
Max--Planck Institut f\"ur extraterrestrische Physik, Postfach 1603, 
D-85740, Garching, Germany\and
Departamento de Astronomia, Universidad de Chile, Casilla 360D, Santiago, Chile
}
\date{Received ......; accepted ......}
\maketitle
\begin{abstract}
 We present observations with the mid--infrared camera ISOCAM on 
board the {\it Infrared Space Observatory} of the major star--forming 
region N\,66 in the Small Magellanic Cloud (SMC) and of its surroundings. 
These observations were performed with broad filters and 
Circular Variable Filters giving a  spectral resolution of about 40.
In addition, CO(2--1) data
are presented, allowing us to identify and study how hot dust relates 
with the different phases of the Interstellar Medium (ISM) present in N66. 
The spectra are dominated by the strong emission of fine-structure line.
Monochromatic maps have been made in the [\NeIII] 15.6 $\mu$m and
[\SIV] 10.5 $\mu$m line. There are significant  differences
between their distributions, due to the effects of density and
of shocks. Aromatic Infrared Bands (AIBs) are seen at various 
places in the field but they are  generally faint.  They
exhibit a variety of shapes and relative intensities, suggesting
that a diversity of carbonaceous materials are present.
Silicate emission is also clearly visible in the central 
condensation and in a few others and emission from hot small grains 
(Very Small Grains, VSGs) longward of 10 $\mu$m is present in the whole region. 
All 
these dust components are heated by the very strong far--UV radiation of the 
many young, massive stars contained in the region. The interstellar radiation 
field (ISRF) at 1600 \AA~ is $\geq$
10$^5$ times the  ISRF of the solar neighborhood in the peaks of 
mid--infrared emission.
The relative contributions of these components (AIB carriers, VSGs and 
silicate grains) to the mid--infrared  spectra seem to
depend on the intensity and the hardness of the far--UV field.
In general the 15/6.75 $\mu$m intensity ratio is higher than in relatively 
quiescent galactic regions (Cesarsky \etal \cite{NGC7023}, Abergel 
\etal \cite{Abergel}) but it is not as high as expected for a linear
increase with ISRF. We interpret this behavior as due to the 
destruction of both AIBs carriers and VSGs in a very high ISRF.
Finally several stars have been detected at 6.75 $\mu$m. Two are red 
supergiants; the other stars are blue and the IR emission is due to 
circumstellar matter or to interstellar matter heated by the star.
\end{abstract}
\section{Introduction}
N\,66 (Henize et al. \cite{Henize}) is the largest and most luminous 
\HII~region in the Small Magellanic Cloud (SMC, m-M=18.94 Laney \& Stobie
 \cite{SMCdist}). It is also known as 
DEM\,S103 (Davies et al. \cite{Davies}) or NGC\,346 referring to the main
 exciting star cluster.

Radio continuum observations   (Taisheng Ye et al. \cite{Ye}) and 
low--resolution H$\alpha$ observations  (Le Coarer et al. \cite{Le Coarer}, 
Fig. 1 and Fig. 5) show that 
the brightest emission region,  N\,66,  is along and to the SW of an 
oblique (SE--NW) ``bar''. 
  A more compact \HII~ region is located 
at $\alpha$(J2000)=00h 59m 16s,
$\delta$(J2000)=-72\degrees 10\arcmin~is N\,66A. A supernova
remnant is located to the East of the region.

A dense cluster of massive young stars is located in N\,66, but there 
are also young stars outside, in particular the ionizing stars of 
N\,66A. Massey et al. (\cite{Massey}) have performed an
extensive study of the stellar content of the region, which contains 
at least 33 O stars, including 11 of type O6.5 or earlier. 22 of these 
O stars are contained in the central star cluster, and the others are 
isolated or in small groups. The hottest star, W\,3 is classified O3 III(f*) 
(Walborn \& Blades ~\cite{Walborn}).
The most massive star W\,1 of the central cluster, classified
O4 III(n)(f) by Walborn \& Blades (\cite{Walborn}),
is in fact multiple and the mass of the brightest component is at 
most 85 \Msun~(Heydari-Malayeri \& Hutsem\'ekers \cite{Heydari}). 
The brightest star in the whole region is HD\,5980, a OB?+WN eclipsing 
binary with
V $\approx$ 11.5 (see for a recent study Heydari-Malayeri et al.
~\cite{Heydari97}); it is located outside the dense cluster.
The region was mapped in the CO(2--1) line with the SEST telescope and 
the results are shown here for the first time.
These data show that N66 does not contain much molecular gas, except for 
a small cloud to the NE of the bar. The whole region has been recently
reobserved in the CO(2--1) line with the SEST at higher sensitivity. These 
new observations show that there is also
weak molecular emission in the N66 bar, connected with the \HII~ region
but not associated with the cloud detected previously  both spatially 
and in the velocity space (Rubio \etal~ \cite{Rubio00}). 
The region is also deficient in \HI~ (Staveley--Smith et al. \cite{smcHI}) and 
the weak 21--cm line emission detected in this area shows no correlation
with the components of N\,66 and may be unrelated.
Probably most of the gas is ionized outside the molecular clouds.
A broad area including the region we studied has been mapped at 68\arcsec~
resolution in the [\CII] 157 $\mu$m line by Isra\"el \& Maloney 
(\cite{Israel}); the intensity peaks on the bar.

In this paper we present mid-IR spectrophotometric observations of N\,66 
obtained with the 32$\times$32
pixel ISOCAM camera on board the {\it Infrared Space Observatory} (ISO) of the
European Space Agency.  These observations belong to a wider program 
aimed to
study the interplay between the interstellar medium (ISM) and star formation
 in our and in external nearby galaxies.  The Magellanic Clouds represented 
obvious sources to include in this project because of their proximity but 
also because they offer the opportunity to study how metallicity influences
 this process. Thus, many HII complexes in both the Large and the Small 
Magellanic Clouds were observed with ISOCAM in spectro-imaging mode between
 5 and 18 $\mu$m and N66 is one of these sources.  
A 6\arcmin$\times$6\arcmin ~field centered on N66 was mapped in 7 
broad--band filters    and the central 3\arcmin$\times$3\arcmin
~have been observed with the Circular Variable Filters (CVFs) as 
dispersive elements. These observations provide a wealth of data on 
warm dust, fine--structure line and Aromatic Infrared Bands (AIBs).

The present paper describes the observations, their 
reduction (Sect. 2) and the analysis of the whole region. We will investigate 
the distribution of the  fine--structure line emission  in Sect. 3, of the 
dust emission of both discrete peaks and  diffuse regions  in Sec 4 and 5. 
Sect. 6 gives the conclusions.  Appendix 1 describes how we built the ISRF 
map at 1600 $\AA$ and
 Appendix 2 contains
a short discussion of the stars seen in our observations.

\section{Observations}
\subsection{ISOCAM observations and data reduction}
The observations were obtained with ISOCAM in September 1996 using a 
6$\arcsec$$\times$6$\arcsec$ pixel field of view for the 32$\times$32 element
mid-infrared camera, covering a 
3$\arcmin$ field of view for each array pointing (see
Cesarsky et al. \cite{CCesarsky} for a complete description).
The observations made with the broad filters were performed as square
3$\times$3--step raster maps with a shift ({\it i.e.} overlap) of 16 
pixels between successive positions giving a final total field of view of 
7.8$\arcmin$$\times$7.8$\arcmin$.
The integration time was 2.1 seconds (time for one exposure). A
number of exposures varying from 7 to 17 according to the filter
were eliminated at the beginning of each raster map. An additional 15 exposures
were taken for each raster position for all the filters in order to insure 
better stability of the detectors, which carry
remanents of their previous illumination history. The  
integration times were
thus 283.5 seconds per filter. The filters were LW2 (5.0--8.0 $\mu$m),
LW3 (12.0--18.0 $\mu$m), LW4 (5.5--6.5 $\mu$m), LW6 (7.0--8.5 $\mu$m),
LW7 (8.5--10.7 $\mu$m), LW8 (10.7--12.0 $\mu$m) and LW10 
(8.0--15.0 $\mu$m).  The raw data 
were then processed in the usual way
using the CIA software\footnote{CIA is a joint development by the ESA
Astrophysics Division and the ISOCAM Consortium led by the ISOCAM PI,
C. Cesarsky, Direction des Sciences de la Mati\`ere, C.E.A., France}. 
A library dark current was subtracted form the broad filter data and a flat--field 
correction was made with a flat field constructed from the data themselves.
The new transient correction described by Coulais \& Abergel 
(\cite{Coulais}) does not give reliable results for the bright point--like
sources. We thus treated the broad band 
ISOCAM images with the software built by Starck \etal~(\cite{Starck}, 
inversion method).
Corrections for field distortion have been applied to filter images
before combining them in each raster. A second--order
flat--field correction was finally used to match the levels on 
contiguous edges of the elementary maps of the rasters. This correction 
is only of a few percent and affects the photometry in a negligible way.
The zodiacal light background used for the raster maps was   the 
lowest emission level in each broad band image. This eliminates efficiently
the zodiacal light which is distributed uniformly, but not if very
extended emission is present.

Full scans of the two CVFs in
the long--wavelength channel of the camera have been performed by decreasing
wavelengths in April 1997. The total covered wavelength range was 5.15 to
16.5 $\mu$m. Each wavelength was observed 12 times in each scan leg, 
with an elementary integration time per measurement of 2.1 second.
The total observing time was 4500 seconds, almost entirely used on--source. 
To correct the raw data for the 
dark current we used the dark model developed  by Biviano \etal~
(\cite{Biviano}) that takes into account the
variation of the dark current inside a single revolution and among
all the revolutions.
Then, the data was deglitched and corrected for the transient response
of the detector. Once again we applied two transient methods, the inversion 
method (Starck \etal, \cite{Starck}) and that which uses the Fouks--Schubert 
equations (Coulais \& Abergel \cite{Coulais}). The latter method gives 
unreliable results in the short channel of the CVF, with negative  flux for 
most of the pixels (before zodiacal light subtraction). This is probably due to 
an overestimation of the dark current which should also be corrected for 
the transient response.
Thus to be coherent with the adopted raster map transient correction, we 
present here the CVF observations corrected with the inversion method.  Flat 
fielding was done using dedicated CVF zodiacal measurements that take into 
account the stray--light due to the mirror and reflections between the CVFs 
and the detector (Boulanger, private communication).

The background of the maps is dominated by zodiacal emission that must
 be subtracted. The whole field of the CVF observations also collects  
extended emission of the N66 region,  preventing us from
measuring  and subtracting the zodiacal light using the classical methods.
In order to correct for zodiacal emission, we used the redundancy between the
raster maps and the CVF observations in the following way:\\
\begin{enumerate}
\item 
We verified that the background emission of the raster maps is 
consistent with the COBE--DIRBE zodiacal light measurements scaled 
for the single orbit in which all the maps were made (Reach, private
 communication). Due to the small size 
of the observed field and to the high ecliptic latitude 
we can safely assume that the zodiacal 
emission is uniform over the field.\\
\item
We measured in the raster maps, using Point--Spread--Function (PSF) fitting, 
the fluxes 
of point--like sources that are also seen with the CVF, and the 
backgrounds around them. For all these sources, we then obtained a measure 
of the sum of the zodiacal background and of the diffuse SMC background in 
the 7 broad--band filters.\\
\item
We built the equivalent of the broad--band filter images 
from the CVF data cube using the transmission curves of the filters 
given in the ISOCAM cookbook, and 
made the measurements of fluxes and backgrounds for the same point sources. 
The source fluxes provide a mutual calibration between
CVF and filters. For each filter the background measurements are  the sum of 
the diffuse emission (N66 and SMC) and of the zodiacal light.
Between the two sets of observations (broad band filter and CVF) the only 
component that could change is the zodiacal light, since 
the solar elongation of the field changed between the filter and 
CVF observations. Thus, the ratio between the two background levels gives the 
variation of the zodiacal light between the two sets of data. We have found 
this ratio to be quite constant, equal to   1$\pm$0.3 for each filter except
for the two shortest--wavelengths filters for which 
the zodiacal emission is very low and our determination uncertain.
We have thus obtained a spectrum of the zodiacal light which is
approximately the same for the filters and the CVF observations.\\
\item The intensity of this spectrum is approximately half of that
of the  CVF zodiacal light spectrum
published by Reach \etal~ (1996), due to the high latitude
of the SMC. We have thus subtracted this spectrum multiplied by 0.5 
from our CVF data.
The final result is a position--wavelength data cube with zodiacal 
light subtracted
from which one can extract spectra at given positions or monochromatic maps.
\end{enumerate}

 There is a slight position shift 
between the observations made with the
Short--Wavelength CVF ($\lambda \leq 9 \mu$m) and those made with the
Long--Wavelength CVF ($\lambda \geq 9 \mu$m) due to a slightly displacement of 
the SW--CVF and LW--CVF. Its effect is unimportant
outside regions like the brightest
emission peaks. In these cases  we sometimes  had to interpolate pixels or
use the spectrum of the Short--Wavelength CVF of the adjacent pixel that 
matched the level of the Long--Wavelength CVF
at 9 $\mu$m in order to produce a reasonable spectrum.

It should be emphasized that the reduction of ISOCAM data is not yet in
its final stage.  We estimate that the uncertainties in the intensities 
of the ISOCAM data presented here are $\simeq$30 \%. 
The main source of uncertainty for both broad band and CVF data  is the
 not complete correction of the detector transient response. For the CVF data
an additional uncertainty arises from  reflections between the CVF and the
 detector (Okumura 2000).

Finally, the 
coordinates given by
the satellite were affected by errors of the order of 10\arcsec~
both for filter and CVF observations. This is due to the lens jitter 
($\simeq$1--2 pixels) and not to the satellite. Fortunately several stars are 
visible
in the filter observations, and allowed  the recentering of the  images on the
Digital Sky Survey (DSS) images (Fig. 5). The CVF frames were
recentered on the filter maps using the star at 0h 59m 27s, 
-72\degrees~09$\arcmin$ 55$\arcsec$ (marked as 755 on Fig. 2 and 5) 
which is detected in 
the 5 $\mu$m continuum map built from the CVF (Fig. 2).  
Figure 1 shows the general outline of the filter observations as a
map in the LW2 filter (6.75 $\mu$m) superimposed on an H$\alpha$ image.

\begin{figure}
\caption{Map of N\,66 in the LW2 filter (6.75 $\mu$m) (grey scale) 
superimposed on an H$\alpha$ image (contours) from the survey of 
Le Coarer et al. (1993) kindly communicated by Margarita Rosado. 
Coordinates are J2000. The LW2 image has been smoothed to
the resolution of the H$\alpha$ one, 9\arcsec. The SE--NW bar is
more marked in the IR image than in H$\alpha$. Notice the spur in 
mid--IR emission extending to the NE of the bar, with no clear 
correspondence in H$\alpha$.}
\end{figure}

\subsection{CO observations and data reduction}

Observations of the CO(2--1) emission line at 230 GHz were obtained as part
of the ESO-SEST Key program: CO in the Magellanic Clouds. The SEST 
telescope, located at La Silla Observatory (Chile) has a 15m diameter and
a FWHM beam at 230GHz of 22". The backend used for the observations 
was a 2000 channel acousto-optical spectrometer (AOS) with a total 
bandwidth of 86 MHz and a channel width of 0.043 MHz. At the frequency 
of the $^{12}$CO(2--1) line the velocity range is 112 ~km~s$^{-1}$
and the velocity resolution is 0.056 ~km~s$^{-1}$.  The observations 
were done in the position--switch mode, with a reference position far from 
the known CO emission zones.  The receiver was a SIS mixer with
system temperature of about 500 K. Intensity values were calibrated  
using the chopper wheel technique. The intensities are given in $T_A^*$
(Kutner \& Ulich \cite{Kutner}) and they take into account the  correction 
for atmospheric attenuation and rearward spillover. To convert $T_A^*$ to 
main--beam temperatures T$_{mb}$ one has to 
divide $T_A^*$ by $\eta$=0.60 at 230 GHz.
Pointing was checked periodically on the SiO maser R\,Dor, and 
a calibration CO spectra towards the SMC source LIRS\,49 (Rubio et al. 
\cite{Rubio}) was taken every day since R\,Dor could only be observed after 
the SMC.

Initially, N\,66 was observed in the CO(1-0) emission line in a coarse grid 
with a 40\arcsec$\times$40\arcsec~ spacing centered 
at $\alpha$(J2000)= 0h 59m 07.5s, $\delta$(J2000)= -72\degrees  10$\arcmin$ 
26$\arcsec$. This grid includes the entire
field studied with ISO. Emission was found at the offset 
position (80\arcsec, -80\arcsec) and a fully--sampled map was done in the 
CO(1--0) line around this region (see Rubio et al. \cite{Rubio}).  
To improve the spatial resolution and derive the 
physical properties of the molecular cloud, N\,66 was later 
fully mapped at 10\arcsec~ spacings, about half the HPBW of the SEST, in the 
CO(2--1) emission line. A 10 $\times$ 11 grid was made, each position observed 
with an integration time of 240 seconds giving an r.m.s. noise of 0.1K per 
channel. The 230 GHz observations were made with half of the high resolution 
backend, the other half being connected to the 115 GHz receiver. The spectra 
were smoothed and a linear baseline was removed.  

A contour map of the CO(2--1) emission integrated over the velocity interval
from 154 to 166 ~km~s$^{-1}$ is shown on Fig. 12, superimposed 
on the LW2 (5.0--8.0 $\mu$m) image. More recently, we discovered faint CO emission
associated with the peaks shown on Fig. 5. This will be discussed in a further
paper (Rubio \etal~ \cite{Rubio00}.

\section{The fine--structure line}
The two fine--structure line [\NeIII] 15.6 $\mu$m and [\SIV] 10.5 
$\mu$m are prominent in the CVF spectra (see Fig. 6). The [\NeII] line at 
12.8 $\mu$m is blended with the infared band (IR) at 12.7 $\mu$m, and this blend is
 quite faint everywhere with respect to the [\NeIII] line. However, 
the distribution of the 12.7 IB + [NeII] emission line (not shown here) and
 that of the [\NeIII] are very different, the former being very similar to the 
distribution of the 11.3 and 6.2 $\mu$m IBs. This suggests that the emission
 feature at 12.7--12.8  $\mu$m is dominated by the IB and that the [NeII] fine 
structure line is everywhere negligible. A strict lower
limit for the [\NeIII]/[\NeII] intensity 
ratio is 1.0. We can compare this lower limit to the [\NeIII] 15.5$\mu$m/[\NeII]
12.8 $\mu$m
 intensity ratios found in other sources with ISO-SWS. Contrary to 
the ISOCAM instrument, SWS  can  separate the [\NeII] 12.8 $\mu$m  emission line 
from the 12.7$\mu$m IB.  The values found are all less than in N66, ranging from 
$\sim$0.8 in the overlapping region of the Antennae galaxy to $\sim$0.2 in the 
Galactic Center (Moorwood \etal~ \cite{Moorwood}, Lutz \etal~ \cite{Lutz}, Kunze
 \etal~ \cite{Kunze}). On the other hand,  higher 
[\NeIII]/[\NeII] values  have been observed with ISOCAM in a sample 
of dwarf galaxies from Madden (\cite{Madden}).
The exceptionally large strength of the [\SIV] line and 
the very high 
[\NeIII]/[\NeII] ratio confirm that
N\,66 is a region of particularly high excitation, due to the large number of 
extremely
hot O stars that it contains (Massey et al. \cite{Massey}).
This was previously known from the
high electron temperatures (12000--14000 K) and the high ratio
[\OIII]$\lambda$5007/H$\beta$ $\approx$ 5 measured at different points of the 
nebula (Peimbert \& Torres--Peimbert
\cite{Peimbert}, Dufour \& Harlow \cite{Dufour}, Pagel et al. \cite{Pagel78},
Dufour et al. \cite{Dufour82}).
 Fig. 3 is a map in the [\SIV] 10.5 $\mu$m line
superimposed on the [\NeIII] 15.6 $\mu$m map.

\begin{figure}
\caption{CVF map of N\,66 in the 5 $\mu$m continuum (contours)
superimposed on the DSS image (grey). Coordinates are J2000. Star N\,346-755 
in the catalog of Massey \etal~ (1989) is marked. It has been used to recenter 
the CVF images on the DSS image.
}
\end{figure}

\begin{figure}
\caption{CVF map of N \,66 in the [\SIV] 10.5 $\mu$m line (contours)
superimposed on the [\NeIII] 15.6 $\mu$m line map. Coordinates are J2000.
Notice the holes in the [\NeIII] 15.6 $\mu$m map. There are large differences
between the distributions of the two line. The position of star W\,3, 
classified O3 III(f*), is indicated by a white cross.
The strong [\SIV] emission to the East is presumably due to
the supernova remnant SNR 0057-7226, whose center is indicated by a black
cross. The solid line represents the cut along which we evaluated the
line intensity profiles of Figure. 4.
}
\end{figure}

\begin{figure}
\caption{The logarithmic profile in the [\NeIII] 15.6 $\mu$m line 
(dashed line) 
and the [\SIV] 10.5 $\mu$m line (solid line) along the cut shown in 
Figure 3. Coordinates are J2000. Note that the [\SIV] line intensity 
is always higher than that of  [\NeIII], the [\SIV]/[\NeIII] line intensity
ratio reaching a maximum at a position closest to SNR.
}
\end{figure}

\begin{figure}
\caption{Map of N\,66 in the LW2 filter
centered at 6.75 $\mu$m (contours), superimposed
on the ESO Digital Sky Survey (DSS) image. Coordinates are J2000.
Several stars are detected
in the LW2 filter: they are HD\,5980 (N\,346-755) and 2 red stars 
(N\,346-283 and 811: numbers in the catalogue of Massey et al. (1989), see
Table 1). The red star N\,66-136 is not detected. 
Some other stars are surrounded by an extended emission: they are 
blue and probably heat the surrounding interstellar matter.
}
\end{figure}

By comparison with the H$\alpha$ map of Fig. 1 and the radio continuum
map of Taisheng Ye et al. (\cite{Ye}), it appears that the fine structure 
line emission is associated with N\,66, extending over our  field of 
view.  We estimate
from the radio data of Taisheng Ye et al. (\cite{Ye}) that the 
amount of energy in the H$\beta$ line is
$\approx$ 5 10$^{-13}$ W m$^{-2}$ in this field. This of course is 
intrinsically corrected for interstellar extinction.
Extinction is low (E(B-V)=0.14 according to Massey et al.
(\cite{Massey})), and we neglect it when considering the mid-IR 
observations.

An interesting feature of the [\NeIII] map is the presence of several
holes in the distribution of the ionized gas (Fig. 3). These holes are 
presumably due to previous supernova explosions or to the effects of stellar 
winds. The latter explanation is probably true for the  central and more 
pronounced hole, which is near the hottest star in the N\,66 OB association 
(an OIII(f*) star marked on Fig. 3).
The differences in the  distribution of the [\NeIII] and [\SIV] line are 
noteworthy. The emission in the [\NeIII] line follows roughly that
of H$\alpha$ as far as one can judge given the different angular 
resolutions (compare Fig. 1 and 3). This is not the case for the 
[\SIV] line. Most of the differences in the main emitting region are 
probably density effects: model calculations e.g. by Stasi\'nska 
(\cite{Stasinska84}) show that the [\NeIII]/[\SIV] line intensity 
ratio is decreased in regions of lower densities, the other 
parameters being the same. This might explain why the holes are more visible 
in the [\NeIII] than in the [\SIV] line.
There is relatively less [\NeIII] emission
in the eastern part of the field where the emission of [\SIV] is
substantial; here the [\SIV] line intensity is roughly twice that of  [\NeIII].
  This region contains a faint H$\alpha$ filament (not visible on
Fig. 1) which is a part of the supernova remnant SNR 0057-7226
(Taisheng Ye et al. \cite{Ye}). The position of the center of this
remnant is indicated on Fig. 3.
The difference in the distributions of the [\NeIII] and of the [\SIV] line
is best illustrated by Fig. 4 which shows the fine structure line--intensity 
profiles along the direction  marked on Fig. 3. It appears that the 
emission of [\SIV] is enhanced compared to that of [\NeIII]
by preferential shock ionization of S with respect to Ne:
in  high-excitation conditions \NeIII~ is the dominant Ne ion and its 
abundance can only be decreased by collisional ionization, while 
\SIII~ and \SIV~ have roughly the same abundances and \SIII~ 
will be ionized in the shock.

By integrating over the whole map, we obtain total fluxes of
$\approx$ 7.3 10$^{-14}$ W m$^{-2}$ and $\approx$ 8.8 10$^{-14}$ W m$^{-2}$
in the [\NeIII] and [\SIV] line respectively. The [\NeIII]/H$\beta$ and 
the [\SIV]/[\NeIII] ratios are $\approx$ 0.1 and $\approx$1.2 respectively
 and together with the optical line ratios obtained by various authors
cited above, they   can be compared with the results of photoionization 
models. We used the models of
Stasi\'nska (\cite{Stasinska82}, \cite{Stasinska84}, \cite{Stasinska90}),
of Stasi\'nska \& Leitherer (\cite{SL96}) and of Schaerer \& de Koter
(\cite{SdK97}). A fair agreement can be reached for the relative
intensities of the [\NeII], the [\NeIII] and the [\SIV] line as well as
of the [\OIII]$\lambda\lambda$5007, 4959 \AA ~and 4363 \AA ~and other
visible line, using models with T$_{eff}$ of exciting stars 40 000 
to 45 000K,
abundances 1/10 solar, density 10 to 100 electrons cm$^{-3}$. These
parameters are reasonable from what we know otherwise of the 
\HII~ region with its very hot exciting stars. Direct abundance effects
cannot be invoked to explain the particularly high [\SIV]/[\NeIII] ratios,
the abundance ratios [Ne/O] (Pagel et al \cite{Pagel78}) and [S/O] (Dennefeld 
\& Stasi\'nska \cite{Dennefeld}) being approximately the same in the \HII~ 
regions of the SMC and of the Galaxy.
However, most of these models yield ratios of the mid--IR line of 
[\NeIII] and [\SIV] to H$\beta$
too high by a factor 2 with respect to  the observed 
ratios, with the exception of the old models of
Stasi\'nska (\cite{Stasinska82}), which are based on the model atmospheres
of Mihalas. But one should note that the optical measurements refer to
the central \HII~ region while the [\NeIII] and [\SIV] line intensity
ratios to H$\beta$ are global. Clearly more detailed optical studies are 
required in order to reach more definitive conclusions.
Given the lack of data for the supernova remnant, it is premature to try to
model the intensities of the [\NeIII] and [\SIV] line in its direction.

\section{The mid-IR emission of the discrete peaks}

The CVF and filter observations show strong emission peaks which we 
discuss here. They are ordered by increasing right ascension and named as
shown in Fig. 5. This figure shows the LW2 (6.75 $\mu$m) contours 
superimposed to the Digital Sky Survey image of N66. The isolated stars 
are identified by  numbers given in Massey \etal~ 
({\cite{Massey}). In Fig. 6 we present the CVF spectra of these peaks.
 Most of the spectra  represent an average of two pixels: spectra of peaks C
 and E have been obtained averaging  four pixels (1 pixel$\approx$1.2 pc for the
 assumed SMC distance). 
In general the spectra show emission bands and fine structure line on top of a continuum.
The wavelengths of the emission bands   correspond to those of the Unidentified
 Infrared Bands already observed before ISO at 6.2, 7.7, 8.6, 11.3 and 12.7 
$\mu$m (Gillett, Forrest and Merrill \cite{Gillett},  Russell, Soifer and
Merrill  \cite{Russella}, Russell, Soifer and Willner \cite{Russellb},
Cohen, Tielens and Allamandola  \cite{cohen85}, Cohen and Kevin \cite{cohen89},
Jourdain  de Muizon \etal~ \cite{Muizon},
Phillips, Airken and Roche  \cite{phillips}, Roche, Aitken and Smith  \cite{roche}).
 They are an universal signature of the ISM 
  in our (Roelfsema \etal~ \cite{Roelfsema}, Verstraete \etal~, 1996, 
Cesarky \etal~ \cite{M17}, \cite{NGC7023}, Boulanger \etal~ \cite{Boulanger},
 Mattila \etal~ 1996, Uchida, Sellgren and Werner \cite{Uchida}) and in 
external galaxies (Boulade \etal~ \cite{Boulade}, Vigroux \etal~ 
\cite{Vigroux96}, Acosata--Pulido \etal~ \cite{Acosta}, Metcalfe etal 
\cite{Metcalfe}, Helou \etal~ \cite{PHOTS}). The exact chemical species 
from which these bands originate  have not been 
yet identified. The   best candidates are the Polycyclic Aromatic
 Hydrocarbons (PAH) (Puget and L\`eger \cite{Puget}), {\it i.e} planar 
macro--molecules (few hundred atoms) transiently heated by single photon 
absorption. However, whatever is the exact nature of these carriers, 
the bands are certainly due to aromatic compounds. For this reason hereafter 
we will call them Aromatic Infrared Bands (AIBs) carriers. Fig. 6 show the
following characteristics:\\
- Peaks A, B, C, E, H and I are aligned along the ``bar''(Fig. 5). Peak C 
coincides
with the center of the dense star cluster NGC\,346. The other peaks are
at various distances from this cluster and receive less far-UV 
radiation except perhaps Peak E.
The H$\alpha$ and fine--structure emission line in the direction of Peak C
are relatively small, presumably because the gas has been partly 
expelled by stellar winds from the dense central cluster.

- Peaks D, F and G lie outside the ``bar''. F coincides with the 
compact \HII~ region N\,66A.

The CVF spectra of all these emission peaks show [\NeIII] 15.6 $\mu$m and 
[\SIV] 10.5 $\mu$m line emission (Fig. 6).

\begin{figure*}
\caption{CVF spectra of the 9 main emission peaks in the region of N\,66.
The peaks are identified on Fig. 5. These spectra have been corrected
for zodiacal light as explained in Sect. 2. An estimate for the mean ISRF  at 1600 \AA~ 
normalized to the  local ISRF at the same wavelength for each source is given.
 If dust is mixed with the ionized gas, these values should be decreased 
by a factor  2.5 (see text for details).
The main fine--structure
line, the visible H$_2$ line and AIBs (A) are identified in the spectrum of 
Peak I. All spectra show the
[\NeIII] 15.6 $\mu$m and [\SIV] 10.5 $\mu$m line. The AIBs exhibit
a variety of shapes and relative intensities. The broad 10 $\mu$m
silicate band is seen in emission in the spectrum of Peak C and B and 
less obviously of Peak F. 
}
\end{figure*}

Even if emission bands are observed at the typical wavelengths of the 
most intense AIBs (6.2, 7.7, 8.6, 11.3 and 12.8 $\mu$m), these are very 
different in their shape and relative intensities from the AIBs observed 
in the galactic reflection nebulae, to which hereafter we will refer as the 
"classical" AIBs.

Peak A shows a broad AIB at 7.7 $\mu$m, a 11.3 $\mu$m AIB not very
 intense and faint 12.7 (possibly blended with a [\NeII] 
line at 12.8 $\mu$m), 13.5 and 14.5 $\mu$m bands.

Peak B shows very faint AIBs, if any, and a broad silicate emission at 
$\simeq$ 10 $\mu$m.  
Note that there are a few faint
hot stars in Peak A (N\,346-320 and 325), as well as in Peak 
B (N\,346-347, 352, 353 and 357: Massey et al. \cite{Massey}).

Peak C, in the direction of the center of the young star
cluster, has a spectrum very similar to that of Peak B but with a 
stronger continuum. It exhibits only faint AIBs
and a broad 10 $\mu$m silicate band is clearly seen in emission.
The spectrum of Peak C
is discussed in more detail by Contursi et al. (\cite{N66}).

The spectrum of peak D is characterized by broad emission near 8 $\mu$m
where the usual AIBs are partly merged. Note the 
short--wavelength continuum, also seen towards Peaks C and E. 
This region contains at least 3 hot stars
(N\,346-466, 469 and 478) the brightest of which is the evolved or reddened N\,346-466
(V=15.91, U-B=-0.54:, B-V=0.27, Massey et al. \cite{Massey})

Peak E contains the relatively
bright, reddened O8V star N\,346-549 with V=15.26, U-B=-0.96, 
B-V=0.22 (Massey et al. \cite{Massey}). The continuum near 5 $\mu$m is the
strongest in the whole map (see Fig.2). It is too strong to be the 
photospheric emission of the
star, but it can be due at least in part to circumstellar dust or to a red
companion. The most conspicuous feature in the spectrum of Peak E
is a very broad emission feature centered near 7.7 $\mu$m in which 
the usual AIBs are even less identifiable than in the spectrum of peak D.
Both the continuum at 5 $\mu$m and the presence of the broad band at 7.7 $\mu$m
 are characteristics of AGN spectra like that of Centaurus A (Mirabel \etal~  
\cite{Mirabel}). The origin of the 7.7 $\mu$m broad feature  has not yet been 
established: it may be due to coal--like  grains. However, it is not clear 
whether these types of grains normally  exist in the ISM of galaxies and become 
visible only when destruction of classic AIBs carriers occurs, or if they form 
through hard UV photons processing on the classical AIB carriers. 
The 6.2 and 11.3 $\mu$m bands are surprisingly weak. The peculiar
appearance of the 7.7 $\mu$m brad feature and the faintness of the 11.3 $\mu$m band 
might be due to some amount of silicate absorption, but the [\SIV] line
at 10.5 $\mu$m, which should also be affected, does not seem particularly weak.
 Moreover, the presence of a certain amount of silicate absorption cannot explain
the weakness of the 6.2 $\mu$m AIB.
Note also the features at 13.5 and 14.5 $\mu$m which  can   
arise from the out--of--plane C--H bending vibrations on aromatic rings with 
3 and 4 contiguous H atoms ({\it trio} and {\it quarto}).

The spectrum of Peak F (N\,66A) shows probable silicate emission and weak AIBs.
Peak F contains at least 7 hot stars, the brightest of which is the
O5.5V star N\,346-593 with V=14.96, U-B=-1.01, B-V=-0.16
(Massey et al. \cite{Massey}).

Peak G coincides with two hot stars, N\,346-628 and 635 (Massey et al. 
\cite{Massey}). This peak is on the molecular cloud
not associated with the main HII region (Fig.12).
Its spectrum is the closest to the typical Galactic AIB
spectra, e.g. those of NGC\,7023 (Cesarsky et al. \cite{NGC7023}).

Peak H has faint bands and peak I displays intense AIB 
bands. Both show a classical AIB spectrum.
They contain a few faint hot stars, respectively N\,346-640, 641, 
648, 654 and N\,346-696 and 697 and in fact it has a steep
 continuum rising toward long wavelength.
Moreover, Peak I contains the bright late O or early B star 
N\,346-690 with V=15.70, U-B=-0.75, B-V=0.00 (Massey et al. 
\cite{Massey}) and  it has the brightest emission in both CO(1--0) 
and H$_2$ among the MIR peaks (Rubio \etal~ \cite{Rubio00}). 
The column density in this peak, relative to 
the others region, is thus sufficiently high to explain the strength  of AIBs. \newline

As the AIBs are believed to be excited mainly by far--UV photons 
in the hard radiation field of N\,66, we have built a rough
map of the radiation density at 160 nm using the  
stellar photometry  from Massey \etal (1989) (Fig. 7). Details about
 how we built this map are given in Appendix 1. There are two sources
 of uncertainties in this calculation.
1) Extinction has not been taken into account (except for determining the 
intrinsic stellar UV flux). Extinction in N\,66 is known to be very small 
for stars (E(B-V)=0.14, Massey \etal \cite{Massey}) and the Balmer decrement
 value of 3.05 $\pm$ 0.15 
(Ye \etal~ \cite{Ye}) is close to the  unreddened  value of 2.86. If dust is 
mixed with the ionized gas, our values for the UV fluxes are upper limits and 
may be too high by $\sim$ 1 mag. (a factor 2.5). If dust is outside the 
ionized gas regions our values are unaffected. 2) The other uncertainty is due
to errors in the assignment of the stellar spectral types.  
However, changing the luminosity class in the 
most ambiguous cases changes the radiation density by only 30$\%$.

The average values of the ISRF at
1600 \AA~ normalized to the local ISRF (LISRF) at the same wavelength 
(Gondhalekar \etal ~\cite{Gondhalekar}) are indicated in Fig. 6 and they range
from 2 to 9 $\times$ 10$^5$ the LISRF. They correspond 
to the values obtained per DSS pixel (=1.7$\arcsec$) averaged over a circular 
area of 2.8 pc radius (= 5.6 DSS pix with an assumed distance for SMC=61 kpc).
This is the approximate resolution of the ISO data, thus the same aperture 
was used to obtained the LW3, LW2 and the 160 nm fluxes reported later in 
Fig. 14.  Note that if dust is mixed with gas inside the HII region,
 the UV flux values still remain very high, ranging from 5.3 
10$^4$ (peaks A and I) to 2.5 10$^5$ (peak C) times that of the solar
neighborhood. 
In Fig. 6 we have not labeled the ISRF average value of peak G because the 
new CO(2--1) data show that this cloud and probably the "spur"
visible as diffuse emission (see Sect. 5) are not associated with the 
N\,66 bar (Rubio \etal~ in preparation). 

\begin{figure}
\caption{Projected distribution of the radiation field at 160 nm 
in the region of N\,66
(grey scale), superposed on the LW2 (6.75 $\mu$m) contours.
Coordinates are J2000. Gray scale values are in Local ISRF units.
}
\end{figure}
From the collection of CVF spectra that we have just discussed, 
several conclusions can be derived:

- Silicate emission is clearly visible in Peak C and B and more marginally in
Peak F. Interstellar silicate emission has been detected in the
Orion nebula and a few other  \HII~ regions, and must be due to relatively big 
grains (size $>$ 0.01 $\mu$m) heated to $\sim$ 100 K or more, 
since it is only seen when the radiation field is very high
(Cesarsky et al. \cite{SilicateOrion}).

- In three peaks (C, D and E),
there is  clear continuum emission at all the
studied wavelengths down to the shortest one, 5 $\mu$m.  While a part of
this continuum may be associated with the AIBs, it is clear that 
they cannot account for all: classic  Galactic AIB spectra
as those of NGC\,7023 or M\,17 (Cesarsky et al. \cite{NGC7023}; 
Cesarsky et al. \cite{M17}) show a negligible contribution of the AIBs at 
5 $\mu$m. The 5 $\mu$m continuum should then be due mostly to grains rather 
close to the hot stars contained in each of these regions. In 
Peak C relatively small silicate or other grains can be heated
to sufficient temperatures to emit at 5 $\mu$m (see Contursi et al. 
\cite{N66}). The radiation field is lower in peaks E and D. But both peaks 
contain reddened stars, and the emission might be 
due to dust around these stars.

- The AIBs emission shows differences in the studied peaks. The AIB
spectra are sometimes very different from the ``typical'' Galactic 
AIB spectra of e.g. NGC 7023 (Cesarsky et al. \cite{NGC7023}). The 
7.7 $\mu$m band towards Peak E is much broader and the 8.6 $\mu$m band is not
visible, perhaps merged into the 7.7 $\mu$m feature  (but this might be
partly due to silicate absorption). Peak D displays 
an intermediate case. ``New'' features near
13.5 and 14.5 $\mu$m are  visible in several spectra. Although faint, 
these features are likely to be real. Residual from glitches could result in
artificial features only for few pixels. Moreover, the same bands  are also
visible in  other regions like M\,17 and NGC\,7023 (Cesarky \etal~ 
\cite{M17}, Cesarky \etal~ \cite{NGC7023}, Klein \etal~ \cite{Klein}). 
Fig. 6 shows that there is a spectral evolution from peak C, in the center 
of the star cluster, to peaks F, I, D and G, where the AIBs are stronger with
respect to the continuum and more similar to the ``classical'' Galactic AIBs. 
This suggests that UV radiation has a crucial role on the grain processing.
It  destroys the classical AIB carriers, favoring the broader--band
emission of relatively big carbonaceous grains which are heated to
sufficient temperatures (peak E) or of smaller carbonaceous grains heated
transiently by absorption of single photons. There might be transformations
from PAH--like 2--D molecules responsible for the classical AIBs into 3--D 
grains or vice--versa. The broad band emission of peak E might indicate 
phenomena occurring close to the reddened O stars contained in these regions.

- The strongest AIB/continuum ratio and the most ``classical'' AIB 
spectrum is observed towards Peak G to the North
of the CVF map. This peak is at the southern edge of a spur well visible in the
LW2 filter images (5.0--8.0 $\mu$m) of Fig. 5. Since this spur and the 
molecular cloud (Fig. 9) are not associated with the N\,66 star cluster 
(Rubio \etal~ in preparation), grains here are heated by an 
ISRF lower than in the bar, providing a spectrum more similar to those 
observed in relatively quiescent regions of our Galaxy. 
 
\section{The diffuse emission}

The diffuse emission is better studied from the filter maps because of
their higher sensitivity. However the CVF observations are useful in 
the interpretation of the filter observations.

There is no reason to doubt that far from the emission peaks which 
coincide with concentrations of hot stars, most of the radiation at wavelengths
shorter than about 9 $\mu$m is due to AIBs and their associated continuum.
This is already clear for Peak G (see Fig. 5) which is far from
the main far--UV sources even if it contains two 16th--magnitude hot 
stars. Consequently, we believe that the
best view of the distribution of the AIBs is offered by the LW2
(5.0--8.0 $\mu$m) map which encompasses the 6.2 and 7.7 $\mu$m features
(Fig. 5 and 6), although there is some contribution from Very Small 
Grains (VSGs: D\'esert et al. ~\cite{Desert}; Dwek et al. ~\cite{Dwek}) 
in the peaks where the radiation field is very
high (see Fig. 6 and Cesarsky et al. \cite{M17}). The stellar 
contribution in this filter is limited to that of a few red stars 
identified on Fig. 5,
and perhaps to the emission of circumstellar dust around hot stars 
as discussed in the previous Section.
The LW6 (7.0--8.5 $\mu$m) and
LW7 (8.5--10.7 $\mu$m) maps (Fig. 8 and 9) and the LW4 (5.5--6.5 $\mu$m) map 
(not shown) are very
similar to each other and to the LW2 map,
although the NE extension and some stars are more easily  visible on the 
LW2 map which is more sensitive due to the broader passband of this
filter.

The filter maps which include AIBs at longer wavelengths, e.g. the
LW8 (10.7--12.0 $\mu$m, not shown) and LW10 (IRAS filter: 8.0--15.0 $\mu$m,
Fig.16) maps,
are more difficult to interpret because they contain a contribution 
of both AIBs and VSGs.
  
\begin{figure}
\caption{Map of N\,66 in the LW6 (7.0--8.5 $\mu$m) filter (contours)
superimposed on the ESO Digital Sky Survey (DSS) image. Coordinates are
J2000. Compare to the LW2 map (Fig. 1, 5 and 7).
}
\end{figure}

A particularly interesting feature in the LW2 (5.0--8.0 $\mu$m) and
LW6 (7.0--8.5 $\mu$m) maps is the emission spur that
extends to the NE of N\,66A.
This spur is probably dominated by AIB emission. It is barely visible
in filters like LW3 (12.0--18.0 $\mu$m) in which the contribution of 
AIBs is minor (see Fig.10).
Fig. 12 shows a superposition of the CO(2--1) line emission in the 
region of N\,66 over the LW2 map.
The CO emission coincides very well with the spur of AIB emission.
As discussed above, this can be easily explained by emission from the surface
of the molecular cloud bathed by a lower and softer radiation field than 
in the bar of N\,66.    

\begin{figure}
\caption{Map of N\,66 in the LW7 (8.5--10.7 $\mu$m) filter (contours)
superimposed on the DSS image. Coordinates are J2000. 
}
\end{figure}

Figure 10 is the LW3 (12.0--18.0 $\mu$m) map of the N\,66 region.
Although there is some contribution from the [\NeIII] 15.6 $\mu$m line 
and of the [\NeII] line and AIB at
12.7 $\mu$m in the LW3 filter, our CVF spectra show that it can
generally be neglected with respect to the continuum.
This is shown by Fig. 11 on which the CVF image in the
continuum on each side of the [\NeIII] 15.6 $\mu$m line (contours)
is superimposed on the LW3 image (grey scale): the agreement is very
good given the differences in field of view and sensitivity. Thus
the LW3 map in our case represents adequately the emission of the Very Samll
Grains (VSGs).
It is noteworthy that the distribution
in the LW3 map is more extended around the ``bar'' than the LW2
map although the latter is more sensitive (compare Fig. 10 with Fig. 5).
This has rarely been seen before and may indicate VSG emission 
in regions where the AIB carriers have been partly destroyed.
\\

\begin{figure}
\caption{Map of N\,66 in the LW3 (12.0--18.0 $\mu$m) filter 
(contours) superimposed on the ESO Digital Sky Survey (DSS) image. 
Coordinates are J2000. This image show the distribution of the warm Very Small
Grains (VSGs).
The faint ``sources'' 1.5\arcmin~ North and South of the main body of 
emission are ghosts of the main peak (Peak C) due to imperfect 
correction of the transient response of the detector.
}
\end{figure}

\begin{figure}
\caption{Map of N\,66 in the LW3 (12.0--18.0 $\mu$m) filter (grey 
scale) superimposed on a CVF image of the continuum near 15.6 $\mu$m. 
Coordinates are J2000. The
agreement is excellent except for a small position shift between the 
filters and the CVF.
It shows that the LW3 image is dominated by continuum emission except
in the NE extension, for which the contribution of the 12.7 $\mu$m
AIB is strong, 
and to the west of the main Peak C, where the contribution of the 
[\NeIII] emission is important.
}
\end{figure}

\begin{figure}
\caption{CO(2--1) emission of the region of N\,66 obtained with a 
resolution of 22$\arcsec$ (contours) superimposed on the LW2 (5.0--8.0 $\mu$m) 
image, which is dominated by the AIB emission (grey scale). Contour levels are 
from -0.5 ($\sim 3\sigma$) to 7.5 in steps of  0.5 K~km~s$^{-1}$, the 
temperature being $T_A^*$. Coordinates are J2000. The CO emission coincides 
with the NE spur of the LW2 map.
}
\end{figure}

 Fig. 13 presents the ``color'' map of the LW3(12.0--18.0 $\mu$m)/LW2
(5.0--8.0 $\mu$m)
intensity ratio.   For building 
this map, the LW3 data have been convolved with the
LW2 PSF as measured on the LW2 map,
and vice-versa before division; this resulted in a small loss of 
resolving
power but produced approximately similar PSFs after convolution. Then
only the part of the data with a signal to noise ratio larger than 
2 after convolution has been retained in both filters.

Previous observations with ISO (e.g. Cesarsky \etal~ \cite{M17}, 
Contursi \etal~ \cite{Contursi}) have shown
that the VSGs start to emit appreciably near 15
$\mu$m when the ultraviolet radiation field is $\geq$ a few  10$^3$ times the 
LISRF. Under these conditions the VSGs temperatures are high enough for their
spectrum to shift towards  
short wavelengths increasing  the 15/6.75 $\mu$m ratio, This ratio ranges from
0.5 to 0.8  in the LISRF environnement. 
The FUV values obtained in N66 indicate that  the ISRF intensity is well above 
10$^3$ times the LISRF everywhere in the
observed region except in the region of the molecular cloud.
In order to study how the 15/6.75 $\mu$m color ratios relate to the
 UV ISRF we have evaluated the 
15/6.75 $\mu$m ratio of each peak over regions of the same size 
(radius = 2.8 pc), 
and plotted them as a function of the ISRF at 1600 \AA~ integrated 
over the same regions (Fig. 14). As expected, the general trend 
is that the higher the ISRF, the higher is the 15/6.75 $\mu$m 
ratio. One can see that the spur (peak G) has a typical 15/6.75 
$\mu$m "cirrus" value of $\simeq$ 1. The same effect is observed
for the global IR emission properties of galaxies. The
15/6.75 $\mu$m -- 60/100 $\mu$m color--color diagram shows that the global 
mid-IR (15/6.75 $\mu$m) colors 
are $\simeq$ 1 for normal galaxies ("cirrus" value) and become significantly 
greater than 1 for more active galaxies (Vigroux \etal \cite{Vigroux2},
  Dale \etal \cite{Silbermann}). The same behavior is also observed inside three 
nearby galaxies, IC\,10, NGC\,1313 and NGC\,6946 
(Dale \etal~ \cite{Dale}).
 
Surprisingly, the highest value of the the 15/6.75 $\mu$m ratio in Fig. 14 
does not correspond to the highest value of the ISRF, located at the
 center of the star cluster (peak C). The CVF spectrum 
of the region with the largest value of the 15/6.75 $\mu$m ratio is 
shown on Fig.15. Following the interpretation of Cesarsky \etal~
(\cite{M17}) and Contursi \etal~(\cite{Contursi}) we would expect a 
continuum towards 15 $\mu$m steeper than that observed 
in Peak C. Fig. 15 shows that this is not the case. The high 15/6.75 $\mu$m 
value observed is due to the nearly complete absence of AIB carriers 
and of continuum at short wavelengths (which is instead present in peak C).
This dramatically lowers  the flux in the LW2 filter. The LW2 and LW3 fluxes of 
this region are respectively $\simeq$ 8 and $\simeq$ 4 times smaller than 
those of peak C in the same filters. The contribution of the [\NeIII] emission 
line in the LW3 filter is only $\simeq$ 10$\%$. We remark that 
this region is close to the earliest--type star of N66 (OIII(f*)) 
suggesting that here the ISRF is not only very strong but also very 
hard. This results in a complete destruction of the AIB carriers 
and partially also of the smallest VSGs. The VSGs might also be destroyed 
in the other peaks, although to a lesser degree. This 
might explain why even if the ISRF throughout the  N\,66 region is  least 
10$^2$ times that of the \HII~ region N\,4 in the LMC, the 15/6.75 
$\mu$m ratios are similar to those found in N\,4 (Contursi \etal, 
\cite{Contursi}). 

Finally, we show on Fig. 16 a LW10 (8.0--15.0 $\mu$m = IRAS 12 
$\mu$m) filter map superimposed on the DSS image: comparison with
Fig. 5 and 10 demonstrates that this image contains features of 
both maps at 6.75 and 15 $\mu$m although it is closer to the 6.75 
$\mu$m map. While
interesting for comparison with IRAS data, the LW10 image is more
difficult to interpret than the images in some other filters which 
have been presented here.

\begin{figure}
\caption{
The LW3(12.0-18.0 $\mu$m)/LW2(5.0--8.5 $\mu$m) color map 
(grey scale) superimposed on the LW3 map (contours). 
Coordinates are J2000. The peak
of the map is at the same location of the peak at $\simeq$  0h 58m 58s
-72$^{\circ}$ 10$\arcmin$ 32$\arcsec$ in Fig. 11. Ratios at some positions
 are indicated.
} 
\end{figure}

\begin{figure}
\caption{The LW3(12.0-18.0 $\mu$m)/LW2(5.0--8.5 $\mu$m) ratios of 
each peak as a function of the average ISRF at 1600 \AA~ normalized to the 
local value of the ISRF at the same wavelengths in 10$^5$ unit.
 These values have been obtained over the same aperture (2.8 pc of radius)
 for each  peak.} 
 \end{figure}
\begin{figure}
\caption{The CVF spectrum of the region with the largest 
value of the LW3(12.0-18.0 $\mu$m)/LW2(5.0--8.5 $\mu$m) ratio.}
\end{figure}

\begin{figure}
\caption{The LW10 (8.0--15.0 $\mu$m = IRAS 12 $\mu$m) image
superimposed on the DSS image. Coordinates are J2000.
Compare to Fig. 5 (LW2 map), Fig. 11 
and 12 (LW3 map) and Fig. 13 (LW7 map): there is a mixture of the 
features of all these maps in the LW10
image, making the latter more difficult to interpret.
}
\end{figure}
\section{Discussion and conclusions}

We have presented CVF and multi--filter ISO observations of the 
region of N\,66 in the SMC. They reveal
a wide variety of phenomena that are not always easy to interpret.
The following results have been obtained:

   i) Emission in the fine structure line [\NeIII] 15.6 $\mu$m and
[\SIV] 10.5 $\mu$m is present throughout the region. These line
are very strong compared to line from singly--ionized ions like [\NeII],
due to excitation by the very hot stars of N\,66. There are considerable 
differences between the space distributions of the [\NeIII] and [\SIV] line, 
that we attribute to density effects in the photoionized regions and to shock 
excitation in a supernova remnant.

   ii) AIB emission is generally weak but present in many places
of the field. This general weakness, already noted by Sauvage et al.
(\cite{Sauvage}), can be related to the low
carbon abundance in the SMC, which is 14--20 times smaller than in 
our Galaxy
(Pagel \cite{Pagel}; Garnett et al. \cite{Garnett}).  However, the 
mid--IR spectrum of a quiescent region in the SMC (Reach \etal~ \cite{SMCB1}) 
is similar to the galactic ``cirrus'' emission, suggesting that the 
differences observed in N66 are principally produced by the extremely 
high and hard ISRF. Analogous ISOCAM observations of another HII region 
in the SMC (SMCB1), not yet completely reduced, will help us to clarify 
which parameters affect the dust properties.
There is AIB emission
probably coming from the surface of a molecular cloud, like in the 
region of N\,4 in the LMC (Contursi et al. \cite{Contursi}). Most 
of the AIB spectra we have obtained (see examples in Fig. 6) are 
different from the classical
"Galactic" AIB spectra, e.g. those of the reflection nebula NGC 7023
(Cesarsky et al. \cite{NGC7023}). The 7.7 $\mu$m feature is almost 
always broader than in the ISM of our Galaxy and the 8.6 $\mu$m AIB 
is not always visible
(merged with the 7.7 $\mu$m feature?). The 11.3 $\mu$m band can be 
strong with respect to the other AIBs, but it can also be quite 
weak. Also interesting is the fact that the bands at 13.5 and 14.5 $\mu$m seem
stronger than waht is observed in the Galaxy.
A similar wide variety of AIB spectra is seen in compact or 
ultra--compact \HII~ regions (Cesarsky et al. ~\cite{M17}, Roelfsema
 et al. ~\cite{Roelfsema}).

\begin{table}
\caption[]{ The global 15/6.75 $\mu$m ratios integrated over the 
CVF and filter fields of view (3$\arcmin$$\times$3$\arcmin$ 
and 7$\arcmin$$\times$7$\arcmin$) of the N66 (SMC) and N4 (LMC) HII }
regions\[
\begin{array}{p{0.15\linewidth}llll}
\hline
\noalign{\smallskip}
HII Region & apparent~size& 15/6.75 & Physical~size & \\
           &        & ratio   & (pc)    \\
\noalign{\smallskip}
\hline
\noalign{\smallskip}
N66 (SMC) &  3\arcmin \times3\arcmin  & 3.3 & 57  & \\
N4 (LMC)  &  3\arcmin \times3\arcmin  & 1.6   & 48 \\
N66 (SMC) &  7.8\arcmin \times7.8\arcmin & 1.2 & 138 & \\
N4 (LMC)  &  6.4\arcmin \times6.4\arcmin & 1.2 & 96 \\
\noalign{\smallskip}
\hline
\end{array}
\]
\end{table}

\begin{figure}
\caption{The average zodiacal--light subtracted CVF spectrum of N66 
averaged over the whole CVF (3$\arcmin$$\times$3$\arcmin$) field, 
corresponding to a physical size of 53 pc $\times$ 53 pc.}
\end{figure}

This variety is presumably due to the co--existence of several forms 
of AIB carriers, one of which dominates depending on the conditions.
The spectrum in the direction of Peak G (close to the edge of the 
molecular cloud) is not conspicuously different from
the classical Galactic AIB spectrum. This cloud is not associated with 
the N\,66 bar. However the 11.3 $\mu$m band is 
somewhat stronger and overall the spectrum is
very similar to the  spectrum of the molecular cloud 
M\,17N (Henning et al. ~\cite{Henning}). The spectrum of peak E  
is similar to that of very small 3--D carbonaceous grains like 
semi--anthracite, which also reproduces well the
spectra of a few Galactic proto--planetary nebulae (Guillois et al.
~\cite{Guillois}). None of our spectra matches well that
of nanoparticules produced by laser pyrolysis of hydrocarbon (Herlin 
et al.~\cite{Herlin}), at least in the 11-14 $\mu$m spectral range.
Together with other data, our data will allow
the study of  the dependence of the AIB spectra on the far--UV
radiation density and spectral hardness. Maps of the radiation field 
like the one presented in Fig. 7 will be useful  for 
such studies.

    iii) The ISRF in the bar of N66 is at least 10$^5$ times the local ISRF 
at the same wavelength. We have evidence that such strong and hard ISRFs 
are able to significantly destroy AIB carriers and to a lesser extent
 also  the  VSGs.

    iv) Aside from the carbonaceous grains just discussed, our 
observations show continuum emission by silicate grains at
several MIR peaks.  

Our observations also shed light on the evolution of the N\,66 region.
Its general optical and mid--IR morphology (Fig. 5) suggests that star 
formation has arisen in an arc of material compressed by shocks, 
probably caused by previous supernovae explosions. There are many 
examples of similar phenomena in both the SMC and the LMC: HI
bubbles (Staveley--Smith \etal~ \cite{smcHI}, Kim \etal~ \cite{LMCHI}),
secondary star formation on 
the edge of these super bubbles (Parker \etal~ \cite{N11}).
This secondary star formation itself is probably not coeval.
In fact, peak C contains only unredenned OB stars and it is more 
evolved than other peaks. Peaks E, H and I for example, have 
reddened stars suggesting that the surrounding material has not 
been yet spread out.
Following the model proposed by Elmegreen (Elmegreen \cite{Elmegreen95}), 
we suggest that star formation along the N66 bar has taken place 
in a sequential way, starting from the OB stars associated to the 
peak C. The spatial separation between the sub-groups of stars 
associated to the different MIR peaks is $\simeq$ 8--10  pc,
comparable to that predicted by numerical simulations 
(Elmegreen  \cite{Elmegreen95}). Also the nature of dust changes along the bar 
(Fig. 6) probably because it forms
at different times and  in different environments. Some of the 
spectra presented in Fig. 6 show weak S(0)(9.6 $\mu$m) and perhaps
S(3) (7.0 $\mu$m) line of H$_2$ in emission. This indicates that some 
molecular gas is still present in the HII region. New CO(2--1) data
seem also support this scenario and show that the MIR peaks correspond to 
molecular clumps with different velocities (Rubio \etal~ in preparation).

Several processes can explain the variety in the observed AIBs strengths 
and shapes. The AIBs faintness can be ascribed to a significant destruction of 
their carriers, either by the harsh ISRF or by shocks  produced by stellar winds. 
Broader than normal AIBs can arise from different grains excited by the ISRF of
 N66; their emission  can be  generally hidden by the AIBs  where these carriers
 are not significantly destroyed. Other possibilities are that the original 
grain size distribution is modified by photo--processing on grains, grains 
shattering,  or that the grain composition  was originally different due to
 the low metallicity of SMC. However, this last hypothesis seems to be
 discarded from the presence of  classical Galactic AIBs in a quiscent
 region of SMC. Finally, the observed  variation in the MIR 
spectra  can be related to the the fact that the dust formed at different
 times and in different environments according to the idea that in N66 the 
star formation evolved in a sequential way.  

The results obtained from the analysis of ISO observations of nearby
objects (HII regions, PDRs, molecular clouds, etc.)
are useful to understand the dust properties of the more distant galaxies.
It is thus important to give the global properties of these nearby 
regions on scale lengths comparable with the ISOCAM resolution of at 
least moderately distant galaxies. As LW2 and LW3 are by far the most 
used filters for ISOCAM observations of external galaxies, we give in 
Table 1 the 15/6.75 $\mu$m (LW3/LW2) ratio of N\,66
for two different fields of view: 3\arcmin$\times$3\arcmin~ (CVF field of 
view) and 7.8\arcmin$\times$7.8\arcmin~ (broad band images), with their 
corresponding physical sizes. This ratio has been calculated directly on the 
broad--band images. For comparison we also give in Table 1 the 
same results for the HII region N\,4 in the LMC, for which ISOCAM broad--band 
data have been already published by Contursi \etal~(\cite{Contursi}). We also 
show on Fig. 17 the spectrum of N\,66 averaged over the total CVF field of 
view. Note that N\,4 is bathed in a 
ISRF 100 times lower than the N\,66 ISRF. At a scale 
length of $\sim$ 100 pc there are no significant differences in the mid-IR 
colors of these HII regions despite their different ISRF. Moreover, on the 
100 pc scale, the LW3/LW2 ratios of both N\,4 and N\,66 are not very 
different from those found in regions bathed in ISRFs similar to the Local 
one. The ISOCAM resolution of $\sim$7$\arcsec$ corresponds to a 100 pc 
region for an object at a distance of $\sim$ 3 Mpc, that of 
galaxies in the Sculptor group for example.

\section{Appendix: building the 160 nm ISRF map of N\,66}
We first determined the 
emission of the stars of N\,66 at 160 nm from their magnitudes and 
spectral types, using the (160 nm - U) colors tabulated in 
Nandy \etal~(\cite{Nandy}). We considered all the 88 OB stars 
catalogued by Massey \etal~(\cite{Massey}) except a few 
for which a spectral type cannot be assessed. 42 of these stars have 
spectral types
given in Massey \etal~(\cite{Massey}). For the remaining stars we 
evaluated the spectral type from their (B-V) and (U-B) colors and 
from the catalog of Azzopardi \& Vigneau (\cite{Azzopardi}). The 
colors have been corrected for reddening assuming E(B-V)=0.14 for 
the stars in the N\,66 cluster (Massey \etal~ \cite{Massey}) and 
E(B-V)=0.09 for stars catalogued in Azzopardi and Vigneau (\cite{Azzopardi}), 
the latter value being the mean one 
for the field stars in the SMC (Garmany \etal ~\cite{Garmany}).
Once a 160 nm flux was assigned to each star, we calculated the 
radiation density at this wavelength in the region. In order to derive 
the geometry of the OB association, we had to eliminate the contribution of 
the diffuse emission from the optical image. We have thus evaluated 
in the DSS image the emission contribution at five different spatial 
frequencies. The image corresponding to the smallest spatial 
frequency (2$\arcsec$) represents the stars themselves. We fitted 
the stellar density distribution by an elliptical gaussian profile 
for the stars belonging to the OB association. We then 
assumed that the depth of the cluster is equal to the minor axis of 
this profile, and assigned to each star a random depth coordinate 
so that the result fits the chosen radial distribution. For the rest 
of the stars we did not take into account their depth distribution 
which anyway is unknown, and assumed that they are all at the same 
distance. This has little consequence as they are isolated or in
small compact groups.  We then calculated the 160 nm 3--D radiation 
density smoothed in cubes whose projection on the plane of the sky 
correspond to the DSS pixel size (1.7$\arcsec$).

\section{Appendix: stars detected in the N\,66 field}

\begin{table*}
\caption[]{Stars visible in the LW2 image and in Massey et al. 
(\cite{Massey})}
\[
\begin{array}{p{0.15\linewidth}lllllll}
\hline
\noalign{\smallskip}
Star & Spectrum & V & U-B & B-V & 6.75 \mu m & Remarks \\
     &          &   &     &     & (mJy)       &         \\
\noalign{\smallskip}
\hline
\noalign{\smallskip}
N\,346-283 &           & 13.38   & 2.52:: & 1.68::  &  7.0 & 
Red~supergiant \\
N\,346-568 &           & 15.97:  & -0.83  & -0.14   & 14.2 &         \\
N\,346-593 & O5.5V+neb & 14.96   & -1.01  & -0.16   & 18.5 & in~N\,66A 
\\
N\,346-755 & OB?+WN    & 11.52:: & -0.85  & -0.24:: & 16.6 & HD\,5980,~ 
var. \\
N\,346-811 &           & 14.16   & 2.68:: & 1.61::  &  3.1 & Red~ 
supergiant~candidate \\
\noalign{\smallskip}
\hline
\end{array}
\]
\end{table*}

A number of stars are visible and identified in the LW2 (5.0--8.0 
$\mu$m) image displayed
Fig. 5. The fluxes we measure are  given in Table 2 
together
with optical photometry from Massey et al. (\cite{Massey}).
For the two red supergiants N\,346-283 and 811, the flux ratio LW2/V 
(6.75/0.55
$\mu$m) is about 0.4, compared to the ratio of 0.2 obtained for the red
supergiant WOH\,53 in the LMC (Contursi et al. ~\cite{Contursi}). The
difference is not very significant and we cannot say if we observe 
the photospheric emission of a M supergiant at an approximate 
effective
temperature of 3000 K, or a hotter supergiant with some circumstellar
emission. A deeper optical and near--IR study is necessary to solve
this ambiguity.

The mid--IR emission of the hotter stars is much  too strong to be
photospheric, and must be either circumstellar or due to some 
interstellar material heated by the star. All these emissions are 
unresolved by our observations. In the case of star N\,346-593 which is
the brightest of the 7 (or more) exciting stars of N\,66A, the 
mid--IR emission is clearly due to interstellar dust heated by 
the star. In the other cases
the emission might be circumstellar since the stars are apparently 
isolated and in relatively gas--free regions.
The most interesting case (the only one for which we have CVF 
observations) is that of HD 5980. In this direction, the AIBs at 8.6,
11.3 and 13.5 $\mu$m are visible while they are absent in the
surrounding area. This indicates carbon--rich dust around the star. 
It is dubious that this dust is circumstellar, since the star is classified
OB?+WN. It is however a very peculiar object which deserves more 
observations before definitive conclusions can be reached.

\acknowledgements{This research has been supported by the ECOS program
of collaboration between France and Chile under grant C97U03. M.R. 
wishes to acknowledge support from FONDECYT (Chile) grant N.$^0$  1990881}
\end{document}